\documentstyle[aps,twocolumn,psfig]{revtex}

\begin{document}
\draft
\title{Actively contracting bundles of polar filaments}
\author{K.~Kruse$^{1,2}$ and F.~J\"ulicher$^{1}$}
\address{$^1$Institut Curie, Physicochimie, UMR CNRS/IC 168, 26 rue
d'Ulm, 75248 Paris Cedex 05, France}
\address{$^2$Max-Planck-Institut f\"ur Str\"omungsforschung,
Bunsenstr.~10, D-37073 G\"ottingen, Germany}
\address{
\begin{minipage}{5.55in}
\begin{abstract} \hskip 0.15in
We introduce a phenomenological model to study the properties of
bundles of polar filaments which interact via active elements. The
stability of the homogeneous state, the attractors of the dynamics in
the unstable regime and the tensile stress generated in the bundle are
discussed. We find that the interaction of parallel filaments can
induce unstable behavior and is responsible for active contraction and
tension in the bundle.  Interaction between antiparallel filaments
leads to filament sorting.  Our model could apply to simple
contractile structures in cells such as stress fibers.                   
\end{abstract}
\end{minipage}
\vspace*{-0.9cm}
}
\maketitle

Living cells have remarkable mechanical properties. In addition to a
passive response to mechanical stresses, eucaryotic cells are able to
actively change their shapes, to generate motion and
forces~\cite{ablrrw94} and to react to externally imposed mechanical
conditions~\cite{to99}. The cytoskeleton, which is a complex network
of elastic protein filaments such as actin filaments and microtubules,
plays a key role in these processes.  Filaments are rod-like
structures that interact with a large number of specific
proteins~\cite{kreis93}. Examples are cross-linking proteins which
induce formation of a gel-like filament network, and bundling proteins
which lead to filaments aligned in parallel.  Furthermore, motor
proteins of the cytoskeleton are able to use the chemical energy of
the hydrolysis of Adenosinetriphosphate (ATP) to generate forces and
motion along filaments~\cite{ablrrw94,kreis93,howa97,jap97}. Myosins
for example interact with actin filaments while kinesins and dyneins
move along microtubules.  The direction of motion of motor molecules
is determined by the polar structure of the filaments which have two
different ends, one denoted ``plus'', the other ``minus'': a given
type of motors always moves towards the same end~\cite{ablrrw94}.
Small aggregates of motors which contain two or more active domains
can bind at the same time to two filaments and exert relative forces
and motion between them.  In contrast to passive filament solutions
whose rheological properties have been studied in recent
years~\cite{mj97}, filament systems which interact with molecular
motors represent intrinsically active materials which exhibit rich
types of behavior that have recently attracted much interest.  In
particular, the formation of asters and spiral defects, as well as the
shortening of filament bundles has been observed in vitro and in
vivo~\cite{takiguchi91,vsb97,nsml97,sewynsl98}. The self-organization
of motor-filament systems has also been addressed
theoretically~\cite{ns96,sn96,basse99}.

Bundles of actin filaments interacting with myosin motors are
prominent cytoskeletal structures that are involved in many active
phenomena in the cell~\cite{ablrrw94}. For example, within the
sarcomeres of skeletal muscle fibers, they are responsible for muscle
contraction; as stress fibers they produce forces in cells; and as
contractile rings they are important during the final step of cell
division. The contraction of sarcomeres follows from a particular
arrangement of myosin and actin filaments. In contrast, stress fibers
and related structures lack an obvious spatial organization of their
components.  The existence of these simple contractile structures
raises interesting questions: What are the minimal requirements for a
bundle of filaments and active elements to contract?  Do stable steady
states of the bundle exist which generate tension?

In this Letter, we develop a simple phenomenological model for
bundles of polar filaments containing active elements, that allows us
to answer such questions and to estimate the mechanical tension
generated in the bundle.  Furthermore, we perform computer simulations
which demonstrate that the general behaviors described by this model
are indeed obtained for simple realizations of the motor-filament
interactions. Finally, we relate our findings
to biological systems.  

Consider a linear bundle of aligned polar
filaments of equal length $\ell$, distributed along the $x$-axis. The
number densities of filaments with their plus-ends pointing to the
right and to the left and with their center located at position $x$
are denoted $c^+(x)$ and $c^-(x)$, resp. These densities satisfy the
conservation laws
\begin{eqnarray}
\label{eq:partialCPl}
\partial_t c^+ & = & D\partial^2_x c^+ -\partial_x  J^{++} -\partial_x
J^{+-}
\nonumber \\
\label{eq:partialCMi}
\partial_t c^- & = & D\partial^2_x c^- -\partial_x J^{-+} -\partial_x
J^{--} .
\end{eqnarray}
\vspace*{-7mm}
\begin{figure}
\label{fig:schema}
\hspace*{2mm}
\psfig{file=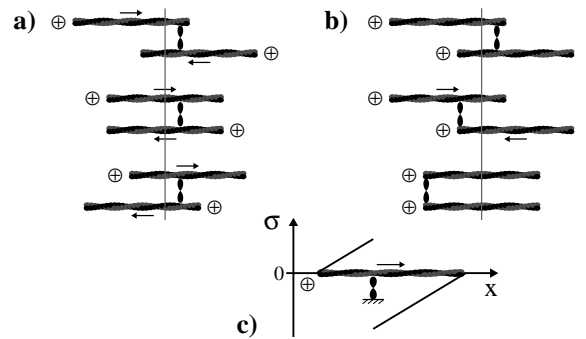,width=7.5cm}
\vspace*{-7.5mm}
\caption{Schematic representation of motor-filament interactions.
Motors are assumed to move towards the plus-end of filaments, arrows
indicate the direction of filament motion.  Vertical lines indicate
the centers of gravity of the filament pair. {(a)} Antiparallel
filaments slide in opposite directions. {(b)} Relative motion of
parallel filaments occurs if a motor binds to a plus-end. Parallel
filaments thus have the tendency to align their plus-ends.
{(c)} Tension profile $\sigma$ along a filament driven by a
point-force.}
\end{figure}
\noindent Here, $D$ is an effective coefficient for filament diffusion
along the  
$x$-axis. The currents $J^{+-}$ and $J^{-+}$ are active filament
currents that result from interactions mediated by motors between
pairs of antiparallel filaments. Currents resulting from interactions
between parallel filaments are denoted $J^{++}$ and $J^{--}$. Here, we
have assumed for simplicity that two-filament interactions dominate,
which corresponds to a sufficently low density of active
elements. Eq.~(1) thus corresponds to a coarse-grained description
where details of the dynamics of the motors have been eliminated. The
direction of filament motion induced by a motor only depends on the
filaments' relative orientation, see Fig.~1. Therefore, we can use
general symmetry arguments to write expressions for the active
currents without referring to a specific interaction mechanism.  To
this end we express the currents in terms of the filament densities
$c^+$ and $c^-$.  For example, we write $J^{+-}(x)=\int_{-\ell}^\ell
d\xi\; j^{+-}(x,\xi)$, where $j^{+-}(x,\xi)$ is the average current of
``minus''-filaments located at $x$ induced by the interaction with
``plus''-filaments located at $x+\xi$. The integral arises since all
filaments with centers located in the interval $[x-\ell, x+\ell]$ can
interact with a filament located at $x$.  Then, assuming that the
probability for an active interaction between two filaments only
depends on the probability of their encounter, we write
$j^{+-}(x,\xi)=v^{+-}(\xi) c^+(x)c^-(x+\xi)$. Here, $v^{+-}(\xi)$ is
the effective relative velocity between two antiparallel filaments a
distance $\xi$ apart induced by many individual events of motor
activity. For parallel filaments we get analogously, e.g.,
$j^{++}=v^{++}(\xi)c^+(x)c^+(x+\xi)$.  Now, symmetry arguments
restrict the possible form of the functions $v^{\pm\pm}$ and
$v^{\pm\mp}$: In the absence of external forces momentum conservation
requires that the center of gravity must remain fixed when two
filaments are displaced. This leads to
$v^{\pm\pm}(\xi)=-v^{\pm\pm}(-\xi)$ and $v^{+-}(\xi)=-v^{-+}(-\xi)$,
see Fig.~1.  Furthermore, invariance of the system with respect to
inversions of space 
\vspace*{-7mm}
\begin{figure}
\label{fig:phasendiagramm}
\hspace*{2mm}
\psfig{file=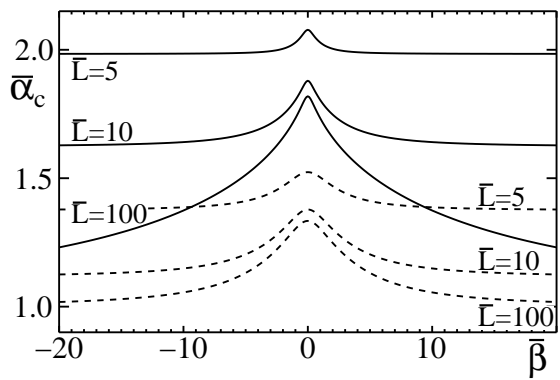,width=7.5cm}
\vspace*{-1mm}
\caption{The critical value $\bar{\alpha}_c=\alpha_c\ell^2c/D$
as a function of $\bar{\beta}=\beta\ell cL/D$ for different values of
$\bar{L}=L/\ell$. Here, $\alpha$ and $\beta$ characterize the
interaction strength of parallel and antiparallel filaments,              
respectively, $L$ is the system size and $\ell$ the filament
length. Solid lines correspond to $\delta c=c/10$ and dashed lines to
$\delta c=c/2$, where $c$ is the filament density and $\delta c/c$ the
relative difference of the densities of plus and minus filaments. The
homogeneous state is unstable for $\alpha>\alpha_c$. }
\end{figure}
\noindent leads to the condition
$v^{++}(\xi)=-v^{--}(-\xi)$. Respecting these 
criteria, we choose for
simplicity $v^{-+}(\xi)=-v^{+-}(\xi)=\beta$ and
$v^{++}(\xi)=v^{--}(\xi)=\alpha\; \textrm{sign}(\xi)$. Here, $\alpha$
and $\beta$ are constants and $\textrm{sign}(\xi)=\pm 1$, depending on
the sign of $\xi$.  We have checked that other choices do not alter
the general properties of our model.  The currents thus read
\begin{eqnarray}
\label{eq:currentPl}
J^{\pm\pm}(x) & = & \alpha\int_{0}^\ell d\xi\;
[c^\pm(x+\xi)-c^{\pm}(x-\xi)]c^\pm(x) \nonumber\\
\label{eq:currentMi}                                     
J^{\pm\mp}(x) & = & \mp\beta\int_{-\ell}^\ell d\xi\; c^\mp(x+\xi)c^\pm(x).
\end{eqnarray}
Equations~(\ref{eq:partialCMi}) and (\ref{eq:currentMi}) describe the
dynamics of our model. We first analyze the stability of the
homogeneous state with constant $c^\pm(x)= c^\pm_0$, which is a
fixed-point of the dynamics for all values of the parameters. Using a
Fourier expansion of the densities $c^{\pm}(x) = c^\pm_0 + \sum_k
c_k^{\pm} e^{ikx}$, the filament dynamics to linear order in the
amplitudes $c_k^{\pm}$ is given by
\begin{equation}
\label{eq:fourier}
\frac{d}{dt}\left( \begin{array}{c}
                    c^+_k\\ c^-_k
                  \end{array} \right) =
 			\left( \begin{array}{cc}
			\Lambda^{++}&\Lambda^{+-}\\
			\Lambda^{-+}&\Lambda^{--}
			\end{array}\right )
                  \left(\begin{array}{c}
                    c^+_k\\ c^-_k
                  \end{array} \right),
\end{equation}
where the elements of the matrix $\Lambda(k)$ are
\begin{eqnarray}
\Lambda^{\pm\pm}(k) & =& -Dk^2-2\alpha(\cos(k\ell)-1)c^\pm_0\pm2i\beta k\ell c^\mp_0 \nonumber \\
\Lambda^{\pm\mp}(k) & =& \pm 2i\beta\sin(k\ell)c^\pm_0 \quad .
\end{eqnarray}
With $\lambda(k)$ denoting the larger of the real parts of the two
eigenvalues of the matrix $\Lambda(k)$, the homogeneous state is
stable if $\lambda(k)\le0$ for all $k$. When an instability occurs, a
band of unstable modes appears, which extends from $k=0$ to some
positive $k$. For a system of size $L$ with periodic boundary
conditions (e.g., a contractile ring), the stability of the
homogeneous state is determined by the sign of $\lambda(k_{\rm min})$,
with $k_{\rm min}=2\pi/L$. We find that the homogeneous state becomes
linearly unstable as soon as $\alpha>\alpha_c$ with the critical value
\begin{equation}
\alpha_c = \frac{D}{\ell^2 c} \;\;f\left(\frac{\beta \ell c L}{D},
\frac{\delta c}{c}, \frac{L}{\ell}\right) \quad .
\end{equation}
Here $c\equiv c_0^+ + c_0^-$, $\delta c\equiv c_0^+-c_0^-$, and
$f(u,v,w)$ is a dimensionless scaling function.  Corresponding         
stability diagrams are displayed in Fig.~2. For all values of the
parameters we find $\alpha_c \geq 0$. Therefore, unstable behavior is
induced by an interaction of parallel filaments with $\alpha>0$.
Consistent with this finding $\alpha_c$ decreases with
increasing $\delta c$, i.e., with an increasing fraction of filaments
pointing in the same direction the system becomes less stable. The
critical value $\alpha_c$ increases with increasing $D$, indicating
that filament diffusion has a stabilizing effect. In many limiting
cases simple analytic expressions for $\alpha_c$ can be obtained. In
particular, we find in the limit of large $L$, $f=2/(1+|\delta c|/c)$
if $\beta=0$ or $\delta c=0$ and $f=1$ otherwise.
In the homogeneous state, the total filament current $J=J^{++} +
J^{+-} + J^{-+}+J^{--}$ vanishes, while currents
of plus and minus filaments exist in opposite 
\begin{figure}
\label{fig:varianzen}
\hspace*{1.5mm}
\psfig{file=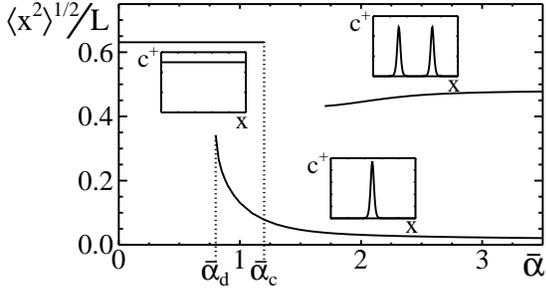,width=7.5cm}
\vspace*{-0.68cm}
\caption{Variance of stable stationary filament distributions
as a function of $\bar{\alpha}=\alpha\ell^2c/D$ obtained by numerical
solution of Eqs.~(1) and (2) for $L=10\ell$ and a system containing
plus-filaments only ($c^-=0$). The homogeneous state is stable for
$\bar{\alpha}<\bar{\alpha}_c$. For $\bar{\alpha}>\bar{\alpha}_d$ a new
attractor corresponding to a shortened bundle appears. A third
attractor is also indicated. The insets show for each attractor
typical filament distributions over one period in arbitrary units.}
\end{figure}
\noindent directions. As soon as
the homogeneous state becomes unstable, a transient current $J$ occurs
which redistributes filaments until an inhomogeneous stationary state
is attained. This final state consists of one or several shortened
filament bundles separated by empty regions.  Fig.~3 displays the
variance $<x^2>=\int dx\; x^2 (c^+ + c^-)$ of stationary filament
distributions as a function of $\alpha$, where we have assumed without
loss of generality that $\int dx \; x(c^+ + c^-)=0$. For small
$\alpha$ only the homogeneous state exists as attractor of the          
system. A second attractor appears for $\alpha=\alpha_d <\alpha_c$,
which corresponds to a shortened filament bundle. A third type of
attractor obtained numerically consists of two shortened filament
bundles. However, we cannot distinguish numerically whether this state
is a real attractor or a long-lived transient state.

The instability of the homogeneous state thus leads to bundle
shortening.  It is triggered by an increase of $\alpha$ and occurs for
all values of $\beta$, which suggests that the interaction of parallel
filaments is responsible for bundle shortening.  On the other hand,
the interaction of antiparallel filaments ($\beta\neq0$) induces
currents which in the unstable regime lead to a separation of plus-
and minus-filaments~\cite{ns96}. The role of interactions between
parallel filaments for bundle shortening can be demonstrated
explicitly for the simple case $D=0$ for which the uniform state is
linearly unstable for all $\alpha>0$. In this case, using Eqs.~(1) and
(2), we can show that for $\beta=0$ and $\alpha>0$ the variance
decreases monotonically, i.e.,  $d/dt \left\langle x^2\right\rangle <
0$, which implies bundle shortening.


Can the active filament interactions which lead to bundle shortening
generate tensile stresses that can be used to perform mechanical work
during shortening? In the bundle, tension arises due to forces                
generated by motors and due to hydrodynamic forces exerted on the
filaments by the surrounding fluid. Let us estimate this tension
$\sigma$ which has units of force. For simplicity, we take only local
friction into account and ignore interactions between different
filaments by friction forces. Consider a rigid rod-like filament
moving in a viscous environment in the direction of its axis with a
friction coefficient per unit length $\eta$. If this motion is induced
by a locally applied force, then the tension profile along the
filament is piecewise linear, see Fig.~1(c). Tension is positive in
the rear part of the filament being ``pulled'' and negative in the
front. Using this tension profile, we calculate an average tension
profile along a filament at position $x$ by summing over all
contributions due to interactions with filaments lying in the interval
$[x-\ell,x+\ell]$. We then calculate the tension in the bundle at
position $y$, by summing the average tension of all filaments with
centers within the interval $[y-\ell/2,y+\ell/2]$. Finally, we coarse
grain over one filament length and obtain
$\sigma(y)=\sigma^{++}(y)+\sigma^{+-}(y)+\sigma^{-+}(y)+\sigma^{--}(y)$, where
\begin{eqnarray}
\sigma^{\pm\pm} & = & \bar\eta \ell \alpha \int \limits_{y-\ell/2}^
{y+\ell/2}dx  \int \limits_{-\ell}^\ell
d\xi c^\pm (x+\xi)c^\pm (x)
\nonumber \\
\sigma^{\pm\mp}& = & \mp\tilde\eta \ell \beta \!\!\! \int\limits_{y-\ell/2}^{y+\ell/2} \!\!\!dx 
 \int \limits_{0}^{\ell} d\xi \left[c^\mp(x+\xi)-c^\mp(x-\xi)
\right]c^\pm (x)
\nonumber
\end{eqnarray}
Here, $\bar\eta$, $\tilde\eta$ are effective frictions per unit length
which result from the coarse-graining and differ from $\eta$ by
dimensionless geometric factors. Note, that the final coarse-grained
tension profile is independent of microscopic details of the
interaction mechanism.

As an example, consider a homogeneous ring, i.e., periodic boundary
conditions. According to the above Equation it generates a tension
$\sigma=2 \alpha\bar \eta\ell^3 ({c_0^+}^2+{c_0^-}^2)$, which for
$\alpha>0$ is positive and could then lead to active contraction of
the ring. This tension persists only if the homogeneous state is
stable. As soon as an instability occurs, i.e., for $\alpha>\alpha_c$,
the ring ruptures and the remaining bundle subsequently shortens. The
shortening of the unstable bundle generates transient inhomogeneous
tension profiles.  Fig.~4 displays instantaneous density and tension
profiles of a filament bundle during shortening.  The density profile
is symmetric, vanishing at the bundle ends.  The tension profile has
the same qualitative shape. The inhomogeneous tension leads to a
filament current $J(x)$ which is antisymmetric corresponding to
filament motion towards the bundle center, consistent with bundle
shortening.

The effective model discussed above does not specify the microscopic
origin of active filament interactions characterized by $\alpha$ and
$\beta$. In particular, the interaction between parallel filaments
could seem surprising~\cite{hk96}. In order to demonstrate that such
interactions can emerge naturally from simple motor-filament interactions we
have performed computer simulations of a more microscopic model.  We
consider $N$ rigid filaments aligned along the $x$-axis.  During each
time-step, a motor complex creates a mobile cross-link between a
randomly chosen pair of filaments. The motors are displaced towards
the plus ends of the filaments with velocity $v$, possibly leading to
a relative filament displacement. This generates an interaction
between antiparallel filaments. A relative displacement between
parallel filaments occurs if we furthermore assume that the filament
ends have different properties than the bulk of the filaments, namely,
a motor which binds to or arrives at the plus end of a filament stays       
attached for some time, see Fig.~1(b). This effect 
generates an effective interaction with $\alpha>0$ between parallel
filaments\cite{bem2}. We have added diffusion steps in the simulations
with $D\neq 0$.  Our simulations show the same qualitative behaviors
discussed above for the phenomenological model: an instability of the
homogeneous state which occurs below a characteristic value of $D$. In
the unstable state we observe bundle shortening due to the interaction
of parallel filaments, interaction of anti-parallel filaments leads to
a separation of plus- and minus-filaments.  Using piecewise linear
tension profiles for moving filaments (corresponding to local
friction), we determine $\sigma(x)$ in our simulation by
averaging over many numerical steps.  For the homogeneous state we
find a tension which fluctuates around a constant positive value. The
simulated density profile, tension profile and averaged filament
current of a shortening bundle are displayed in Fig.~4.

Concluding, we have demonstrated that a bundle of aligned filaments
and motors may contract and generate tension even if it is lacking a
spatial organization. 
Using symmetry arguments, we have shown under
some simplifying assumptions that the interaction of parallel
filaments is important. Whether stress fibers in cells use the
physical mechanism proposed here for contraction remains to be         
tested. In particular, experiments which focus on the interaction
between parallel filaments should be performed. Stress fibers are
likely to be more complex than the system studied here and might also
use other mechanisms for contraction. For example, they could have
some yet unknown spatial organization that allows
\begin{figure}
\label{fig:kontraktion}                                  
\hspace*{1mm}
\psfig{file=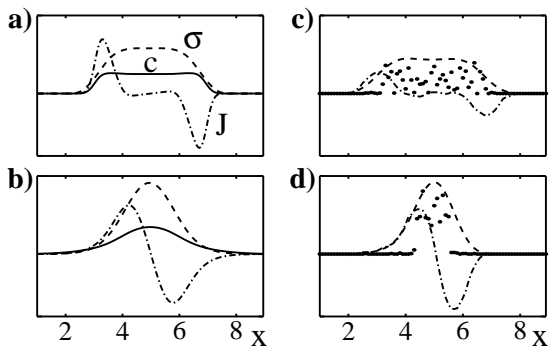,width=7.5cm}
\vspace*{-4mm}
\caption{Instantaneous filament distributions (full lines, dots), the
tension profile (dashed lines), and the current (dash-dotted lines) in
arbitrary
units of contracting filament bundles. {(a)}, {(b)} Numerical
solutions of Eqs.~(1) and (2) with $\bar\alpha=0.5$, $\bar\beta=0$,
$c=0.5$, $\delta c=1$, and $\bar L=10$. {(c)}, {(d)} Simulation of
$N=1000$ filaments as described in the text.
{(a)} and {(c)} are for earlier, {(b)} and {(d)} for later
times.}
\end{figure}
\noindent them to contract
more efficiently, or other components in addition to motors and
filaments might be involved. Furthermore, they might work close to the
percolation threshold where large filament aggregates occur and which
is not captured by our model.
Nevertheless,
we think that our model helps clarifying the role of motor-filament
interactions and self-organization phenomena for force generation
in biological cells.

We thank M. Bornens, S. Camalet, P. Janmey, A. Maggs, F. N\'ed\'elec,
A. Ott, A. Parmeggiani, M. Piel, J. Prost and A.B. Verkhovsky for
stimulating discussions.


\begin{thebibliography}{1}
\bibitem{ablrrw94} B.~Alberts, et al., {\it The molecular biology of
the cell}, 3rd ed., Garland, New York~(1994).
                                                                      
\bibitem{to99} O.~Thoumine and A.~Ott, MRS Bulletin~{\bf 24}~(10), 22
(1999).

\bibitem{kreis93} T.~Kreis and R.~Vale, {\it Cytoskeletal and Motor
Proteins}, Oxford University Press, New York (1993).

\bibitem{howa97} J. Howard, Nature {\bf 389}, 561 (1997).

\bibitem{jap97} F.~J\"ulicher, A.~Ajdari, and J.~Prost,
Rev.~Mod.~Phys.~{\bf 69}, 1269 (1997).

\bibitem{mj97} F.~C.~MacKintosh and P.~A.~Janmey, Curr.~Opin.~Solid
State \& Mat.~Sci.~{\bf 2}, 350 (1997).

\bibitem{takiguchi91} K.~Takiguchi, J.~Biochem.{\bf 109}, 502
(1991).

\bibitem{vsb97} A.~B.~Verkhovsky, T.~M.~Svitkina, and G.~G.~Borisy,
J.~Cell Sci.~{\bf 110}, 1693 (1997).

\bibitem{nsml97} F.~J.~N\'ed\'elec, et al., Nature (London)~{\bf 389},
305 (1997).                                                              

\bibitem{sewynsl98} T.~Surrey, et al., Proc.~Natl.~Acad.~Sci.~USA~{\bf
95}, 4293 (1998).

\bibitem{ns96} H.~Nakazawa and K.~Sekimoto, J.~Phys.~Soc.~Japan~{\bf
65}, 2404 (1996).

\bibitem{sn96} K.~Sekimoto and H.~Nakazawa, in {\it Current Topics in
Physics},
Y.M. Cho, J.B. Hong, and C.N. Yang (eds.) vol1, p.394 (World
Scientific, 1998); physics/0004044.

\bibitem{basse99} B.~Basetti, M. Cosentino Lagomarsino and P. Jona,
cond-mat/9911140 (1999).

\bibitem{hk96} There is however evidence for active interactions of
parallel microtubules, see A.~A.~Hyman and E.~Karsenti, Cell~{\bf 84},
401 (1996).

\bibitem{bem2} In general, an interaction between parallel filaments
is generated if motor function depends on the distance of the motor
from the filament ends. Note, that the sign of $\alpha$ depends on the
details of the mechanism.

\end{thebibliography}
\end{document}